\documentclass[conference]{IEEEtran}
\IEEEoverridecommandlockouts
\usepackage{cite}
\usepackage{amsmath,amssymb,amsfonts}
\usepackage{algorithmic}
\usepackage{graphicx}
\usepackage{textcomp}
\usepackage{xcolor}
\usepackage{booktabs}
\usepackage{svg}
\usepackage{wrapfig}
\usepackage{url}
\usepackage{comment}
\def\BibTeX{{\rm B\kern-.05em{\sc i\kern-.025em b}\kern-.08em
    T\kern-.1667em\lower.7ex\hbox{E}\kern-.125emX}}
\begin{document}

\title{Tracking research software outputs in the UK\\

\thanks{Funding redacted for peer review.}
}

\author{\IEEEauthorblockN{Domhnall Carlin}
\IEEEauthorblockA{\textit{Centre for Secure Information Technologies} \\
\textit{Queen's University Belfast}\\
N. Ireland, UK \\
d.carlin@qub.ac.uk}
\and
\IEEEauthorblockN{Austen Rainer}
\IEEEauthorblockA{\textit{School of Electronics, Electrical Engineering and Computer Science} \\
\textit{Queen's University Belfast}\\
N. Ireland, UK \\
a.rainer@qub.ac.uk}
}

\maketitle

\begin{abstract}
Research software is crucial in the research process and the growth of Open Science underscores the importance of accessing research artifacts, like data and code, raising traceability challenges among outputs. While it is a clear principle that research code, along with other essential outputs, should be recognised as artifacts of the research process, the \textit{how} of this principle remains variable. This study examines where UK academic institutions store and register software as a unique research output, searching the UKRI's Gateway to Research (GtR) metadata for publicly funded research software in the UK. The quantity of software reported as research outcomes remains low in proportion to other categories. Artifact sharing appears low, with one-quarter of the reported software having no links and 45\% having either a missing or erroneous URL. Of the valid URLs, we find the single largest category is Public Commercial Code Repository, with GitHub being the host of 18\% of all publicly funded research software listed. These observations are contrasted with past findings from 2023 and finally, we discuss the lack of artifact sharing in UK research, with resulting implications for the maintenance and evolution of research software. Without dissemination, research software risks demotion to a transient artifact, useful only to meet short term research demands but ultimately lost to the broader enterprise of science.
\end{abstract}

\begin{IEEEkeywords}
Repository, Reproducibility, Research Software, Academia, Software Maintenance
\end{IEEEkeywords}

\section{Introduction}

%%FROM SANER SUB
Institutional repositories (IRs), or Research Information Systems (RIS), maintain permanent records of output from research-focussed employees of the host institution\cite{Carlin2023}. While these IRs originally came into existence to fulfil progressive open access requirements of funding bodies, they now seek to digitally preserve the intellectual output of the university and its researchers \cite{crow2002case}.  
The benefits of Open Science have served to highlight the necessary access requirements to artifacts of the research beyond open access to papers alone, e.g. data and code. However, this creates challenges in the maintenance of traceability between all artifacts and outputs of the research process \cite{hata2021science}. While it is a clear principle that research code, along with other essential outputs, \textit{should} be recognised as artifacts of the research process, the \textit{how} of this principle remains much more variable. Due to the pressure to wrap research software into a research paper, rather than focusing on the \textit{software} as an artifact in and of itself, it can be particularly difficult to map research software.
%CHECK THIS
However, employees should receive credit for their work, which is one of the founding principals of the Research Software Engineering (RSE) movement.  If this academic credit system relies on the IR, how can they get credit, especially if the IR structurally them for getting credit for their outputs or software in this case.  This is exacerbated by the role of software maintenance, where maintaining software might get more citations eventually, but the effort/credit ratio is far below that of publishing a new paper. Therefore, when you stack the credit to the release of new outputs, the RSE in a maintenance role lacks credit, so the motivation to maintain and evolve software dwindles.
%END CHECK

As the FAIR principles and Open Science best practices gain wider adoption and continue to evolve, particularly for software \cite{FAIRrs}, there is a growing need to identify research outputs linked to scientific publications \cite{garijo_2024}. Carlin, Rainer and Wilson \cite{Carlin2023} distinguish between code-centric repositories, replete with tools to aid in active and collaborative software development, (e.g. GitHub, GitLab etc.), and purely archival repositories, i.e. IRs.  %This is mapped in Fig.\ref{fig:repositories}, with some examples of repositories that are archival, but with developer-friendly features.
In their special issue on software citation, indexing and discoverability, Katz and Chue Hong call for \textit{`More consistent use of best practices for registries and repositories that store or refer to software.'}\cite[p5]{katz2024special}. There are valuable data in these repositories, but if they cannot be found readily, reusing the software to reproduce and replicate results is a difficult task.

\begin{comment}
    
\begin{figure}
    \centering
    \includegraphics[trim=1.7cm 3.05cm 1cm 2.65cm, clip, width=1\linewidth]{figs/Slide48.jpg}
    \caption{Examples of repositories on scale from developer-friendly to archival.}
    \label{fig:repositories}
\end{figure}
\end{comment}

\section{Previous work}\label{PrevWork}

There have been increasing efforts to assess the use of code repositories for links to academic papers, and vice-versa.
The use of public code-focused repositories for research software has been found to have grown enormously \cite{10.1007/978-3-031-16802-4_15}.  In an analysis of the use of GitHub, SourceForge, Bitbucket and GitLab between 2007 and 2021, the authors found these repositories were linked to from academic papers in ArXiv and the PMC Open Access Subset 160 times in 2007, rising to 76,746 times in 2021. Further, the authors found that by 2021, one in five papers in ArXiv contained a URI to GitHub.
Escamilla et al. \cite{EscamillaEmily2023CBNA} highlight the cessation of Git Hosting Platform (GHP) services, such as Google Code and Gitorious, as serious issues for reproducibility.  The authors analysed GHP URIs from 2.6 million articles in ArXiv and PubMed Central to examine bitrot in research software.  93.98\% were still accessible on the live Web. They also reported that 68.39\% were captured by Software Heritage, and 81.43\% had at least one archived version in a web archive.
In a sample of the dataset from \cite{WATTANAKRIENGKRAI2022111117}, Hata et al. found at times there was no established link specified between the published paper and a software repository, even though the authors and repository owners were the same people \cite{hata2021science}.  Färber \cite{10.1145/3383583.3398578} mapped all GitHub code repositories linked to in scientific papers, using the (now defunct) Microsoft Academic Graph and found GitHub to be the most commonly reported repository host, with 4,876 repositories, a finding shared by \cite{WATTANAKRIENGKRAI2022111117}. This also aligns with Hasselbring et al \cite{hasselbring2019fairopencomputerscience}, who estimated that, at the time of writing, GitHub contained over 5,000 public repositories of research software.

Garijo et al. \cite{garijo_2024} investigated the bidirectional links between academic research papers and their corresponding code repositories. The authors presented two article-code repository extraction pipelines for identifying these links, using 14,000 PDF and \LaTeX source files from ArXiv's Software Engineering to yield 1,400 such links.
, finding 1400 links in 14,000 manuscripts in ArXiv's Software Engineering section.
Wattanakriengkrai et al. \cite{WATTANAKRIENGKRAI2022111117} investigated the connection between networks of academic publications and GitHub repositories. The study examined 20,000 GitHub repositories that cited academic papers, usually through the README.md file. Over half of these repositories implemented methods or algorithms from others' papers, while 40\% referenced the repository owners' own research publications. To explore the link from papers to repositories, the authors analysed a sample of 2,032 academic papers from seven leading software engineering publication venues. They found that most of these papers did not link to any repository, but when they did, GitHub was typically cited. Notably, none of the papers linked back to an institutional repository, and only five public repositories appeared across the entire dataset.  This lack of inclusion of IRs within such analyses motivated \cite{Carlin2023} to analyse IRs across the UK, inferring issues regarding the recognition of research software as academic output.
Past attempts to survey the Research Software landscape have been hampered by the lack of links within research papers to the underpinning software and vice-versa.  

The present work seeks to establish where software, as a distinct research output, from UK Academic Institutions is kept/recorded/registered by using public records, rather than links to/from manuscripts.  The motivation for this is to gain an understanding of the current levels of research software registration and where such outputs are stored.  As software is fundamental to the research process, it is therefore also a fundamental step in both reproducibility and replicability of that research.
%END FROM SANER SUB

\section{Method}
The UK Research and Innovation (UKRI)'s Gateway to Research (GtR) is a web portal to enable users to search for and analyse metadata about all outputs and outcomes from publicly funded research and innovation in the UK \cite{Ukri2025Feb}. The award data are published quarterly, updating April, July, October and January \cite{Ukri2025Feb}. This data includes all claimed outputs from public funding through UKRI, allowing analysis of outputs since 2006 with rich metadata.  To provide the necessary data for the present work, this dataset was queried in August 2023 and again in February 2025 for all software listed as research outcomes, with no restrictions on date range\footnote{\url{https://gtr.ukri.org/search/outcomes?term=*&selectedFacets=c29mdHdhcmVUeXBlfHNvZnR3YXJlfHN0cmluZw==&fields=&type=softwareandtechnicalproduct&fetchSize=25&page=1&selectedSortableField=score&selectedSortOrder=ASC}}.  A snapshot CSV file of this report was generated, which forms the basis for all further analyses.  All code and data used to generate the results is available at \url{https://anonymous.4open.science/r/outcomes_software_data-FB84}.

%\subsection{URL analysis}
The URL for each entry was examined in two stages.  Firstly, a check was made to determine if an entry exists in the data set (that is, the submitter of the output supplied a URL as part of the output metadata).  If no URL was supplied, the URL was listed as \textit{`missing'}.  Secondly, if a URL was listed for the entry, a check was made using the Python Requests library \cite{python_requests} for the HTTP response status code to test for liveness of the URL.  These were logged along with the URL for analysis. Initially, a large amount of HTTP redirection status code responses were returned (i.e. HTTP 3**).  This appeared to be due to the mass migration of Zenodo URLs to an alternate path URL.  As this redirection would be ordinarily handled by a browser using the \texttt{location} field in the HTTP 301 status code response, explicit redirection was enabled for the automated URL processing.  A brief sub-analysis was conducted with the URLs that returned an error to the Python Requests library (i.e. not an HTTP Status code), to retrieve expiration dates of the domain registration.

%\subsection{URL Categorization}
To determine where research software outputs were stored, each URL was categorized into sets of pre-determined categories, based on key words as established in an early version of the present work %\{se4rse23}
[anonymous]. In that early work, 3000 URLs from the data as at July 2023 were manually categorized to aid the development of either a keyword or a library of categorized URLs to avoid repeated effort. These categories reflect the \textit{type} or \textit{intent} of the storage in which the software listed is found.  If a URL did not fit one of the categories from the previous work, a new category was manually created where necessary.  Approximately 200 URLs needed manual categorization for the present cohort.  The final category set is discussed further in Section \ref{results_section}.

\section{Results}\label{results_section}
%\begin{comment}
    
\subsection{Metadata}
\subsubsection{Counts per Lead Research Organisation}
169 organisations submitted software.  The average count per institute was 53.14, with 44/169 (26\%) exceeding this. More than 50\% of software outputs were from the top 12 organisations ranked by total count of software records. Notably, all 12 are Russell Group universities, a self-selecting group of 24 `research intensive' UK universities.

\subsubsection{Counts per PI} Each Principal Investigator (PI) is assigned a unique reference number to track outputs through GTR.  2597 separate PIs contributed at least one item of software, with the highest count being 144 and a mean of 3.33.
%\end{comment}

\subsection{URL Analysis}

\begin{figure*}
    \centering
    %[trim={left bottom right top},clip]
    \includegraphics[trim={5 100 4 3},clip, width=\textwidth]{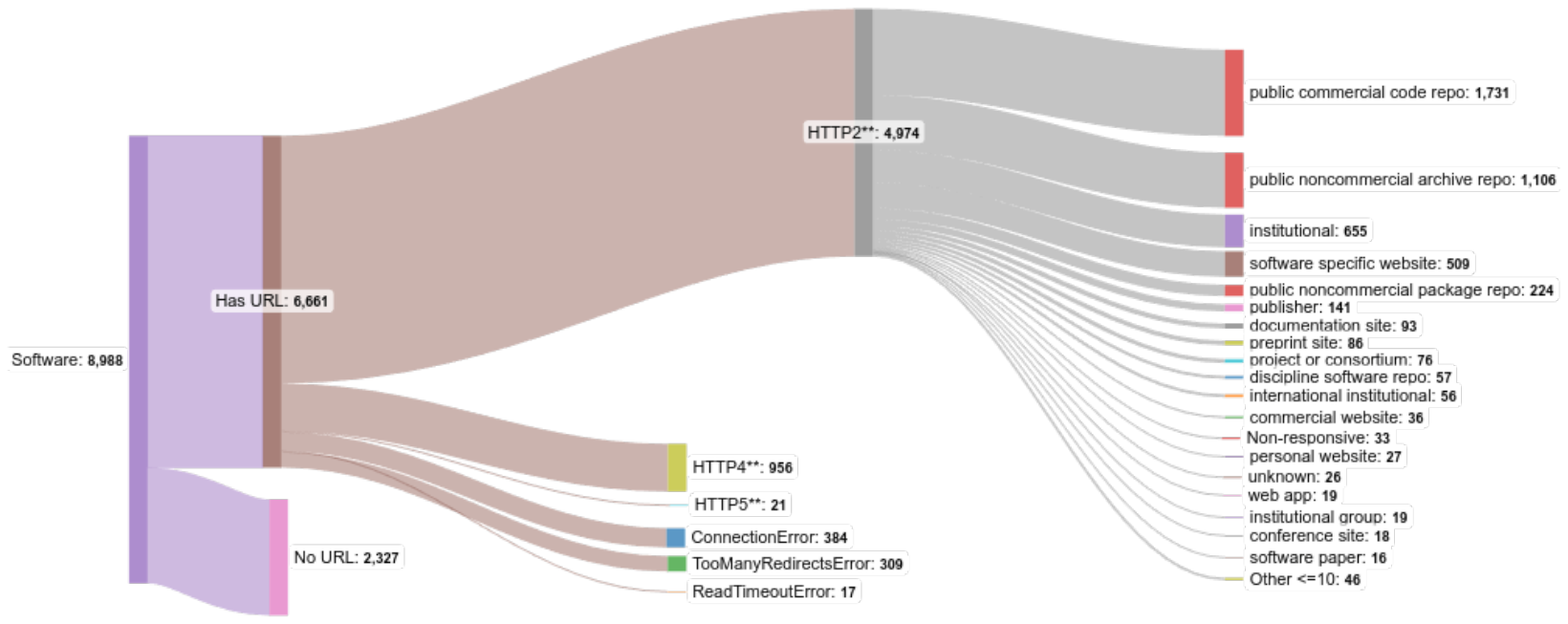}
    %\includesvg[width=SIZE\textwidth]{figs/sankeymatic_20250407_192326_1000x500 (1).png.svg}

    \caption{Numerical flow of data from all software to categories.}
    \label{fig:sankay_software}
%\end{figure}
\end{figure*}

Figure \ref{fig:sankay_software} depicts the numerical data flow from all software entries to each category. Out of 1,056,247 listed \textit{outcomes}, there were a total 8988 entries under the category of \textit{software} within GtR, representing 0.85\%. Of these, 2327 (25.89\%) had a missing URL, with 6661 containing a URL.  Of the 6661 present URLs, 4974 (74.67\%) responded with an HTTP 20* status code, indicating they were successfully resolved.  Table \ref{tab:http_status_distribution} shows the remainder of the HTTP responses.
Of the non-2** responses, HTTP status code 404 (not found) was the most prevalent, accounting for 570 (8.56\%) of all URLs.  HTTP 403 (forbidden) accounted for 295 (4.43\%) of URLs.  This suggests protection mechanisms (e.g. against Cross Site Request Forgery (CSRF)) or services behind network-protection, such as Cloudflare.  For ethical concerns, no upgraded attempt was made for these URLs, which is discussed in Section \ref{section:discussion}.

The largest two categories of working URLs (Public Commercial Code Repo and Public Non-commercial Archive Repo) represent 57\% of the entries.  Sub-analysis showed that GitHub was clearly the most common host, representing 91.45\% of public commercial code repositories. It also represented 24.32\% of all URLs provided and 31.81\% of working URLs.  Within the Public Non-commercial Archive Repo category, Zenodo represented 90.7\% of the category and 21.17\% of all working URLs.
%\begin{comment}
\begin{table}
    \centering
    
        \caption{Distribution of HTTP Status Responses}
    \begin{tabular}{lrr}

        \hline
        \textbf{HTTP Status} & \textbf{Count} & \textbf{\%} \\
        \hline
        2** & 4974 & 74.67\% \\
        4** & 956 & 14.35\% \\
        5** & 21 & 0.32\% \\
        ConnectionError & 384 & 5.76\% \\
        ReadTimeoutError & 17 & 0.26\% \\
        TooManyRedirectsError & 309 & 4.64\% \\
        \hline
        \textbf{Total} & 6661 & 100.00\% \\
        \hline
    \end{tabular}
    
    \label{tab:http_status_distribution}
\end{table}
%\end{wrapfigure}
%\end{comment}

\subsection{Erroneous URLs}
401 erroneous URLs were checked if they had expired according to their WHOIS record.  When duplicate domains were removed, 191 remained.  Of these, only two had expired registration dates, 44 had current registration dates and 145 returned an error (e.g. no record or expiration date was available). 309 URLs generated a TooManyRedirectsError. The Python Requests library defaults to a limit of 30 and this exception is raised when the request exceeds that limit.  It is unlikely that a legitimate response would include 30 redirects, so this may again be a protection mechanism against automated web-scraping.

\subsection{Change over time}
When comparing the current snapshot of GtR to the July 2023 data in [anonymous], there are some noticeable changes.  The number of software outcomes increased by 1756 (24.28\%). The number of URLs supplied increased by 1492 (28.86\%), with the working URL count up by 1031 (26.13\%).  As a percentage of the software outcomes listed, in July 2023 71.47\% had URLs and 54.56\% had working URLs.  In 2025, this has risen to 74.11\% and 55.37\% respectively.

\section{Discussion}\label{section:discussion}
Fundamentally, research software that cannot be found, cannot be reused, maintained or evolved. The key aim of the present work was to assess the number of software outputs of publicly funded research and to establish where software, as a distinct research output, from UK Academic Institutions is reportedly kept/recorded/registered.  There has been much effort in establishing guidelines for FAIR research software (e.g. \cite{FAIRrs}), toolsets to enable citation (e.g. \cite{druskat2017citation}) and funder policies for such artifacts (e.g. \cite{ukri_concordat}). However, the quantities of software reported as research outcomes remain low in proportion to other categories.  Almost half of the reported software had no working URL to enable reuse, which should be mandatory.  This represents a lapse in fundamental open science principals.  Peer-reviewed publications are recognised as the primary output from research and they act as proxy measure of a researcher’s or research group’s contribution to science. These publications are not possible without the research software that powers that research. If we can’t reliably connect (or trace) research software with the respective publications, it becomes difficult, if not impossible, to demonstrate the contribution that research software makes to science. The lack of traceability is therefore hiding, however unintentionally, the value of research software and limiting its reuse. With reproducibility being a key marker for the validity of experimental results \cite{acm_repr}, it is difficult to see how, given current practices evident within this analysis, this is being attained.
%ADD BIT ABOUT POPULARITY OF GITHUB AND ZENODO MAY BE PARTLY CAUSE THEYRE FREE AND DONT REQUIRE MAINTENENCE.
If research software remains unpublished or inaccessible beyond its initial developers or user-base, there can be subsequent consequences for its reuse, maintenance and evolution.  If open source software maintenance is to be considered a socio-technical endeavour \cite{Mens2016}, then failing to share the software excludes the community element, i.e., the external contributors.  These external users and contributors can highlight new or unforeseen issues with evolving dependencies and library ecosystems.  Without such community oversight, the risk of obsolescence, bit rot and technical debt can increase.  Maintenance then remains the sole responsibility of the original developer(s), who typically work(s) in time- and budget-bound blocks.  With software citation being an established concern but with limited uptake, there are few academic incentives for maintaining software, as opposed to disseminating rapid publishable research results.  Coupled with the lack of funding beyond the lifetime of a project, the outcome can be a tool that, while adequate for immediate research needs, prioritises short-term results over long-term solutions.

Beyond the benefits to open science and software evolution, there is a real applied risk to open source research software that is not correctly linked or regularly maintained.  Software supply chain attacks, such as \textit{repojacking}, \textit{dependency confusion} or \textit{typosquatting}, have begun to focus on open source package repositories that are designed to promote friction-reducing installation and are popular among researchers.  Recent examples (e.g., \cite{sharma2022pymafka, kuznetsov2022lofylife, checkpoint2022cloudguard}) have shown the potential impact of such attacks, which are trivially implemented.  Failing to supply a URL that points to research output is not only flouting open science principles, but is potentially increasing the attack surface for open source repositories.

\section{Future Work}
The present work focuses on UK research outputs through GtR, highlighting the main findings of early stage research.  The next phase will examine the growth and decline over time of the various categories and what insights this can provide into trends in academic software publication.  Expanding the base dataset, future research could examine if similar trends are identified in other jurisdictions, e.g. U.S. National Science Foundation's Public Access Initiative, the European Union's CORDIS portal and the Australian Research Data Commons, among others.  This will help build a broader evidence base for understanding and comparing academic software sharing practices globally.

%TODO ethical scraping, timeouts, unknonw- talk about how improved and reducing effort. Find links in papers to gtr and viceversa.

  \bibliographystyle{ieeetr}
  \bibliography{IEEEabrv, Tracking_research_software_outputs_in_the_UK}

\begin{thebibliography}{10}

\bibitem{Carlin2023}
D.~Carlin, A.~Rainer, and D.~Wilson, ``Where is all the research software? an analysis of software in uk academic repositories,'' {\em PeerJ Computer Science}, vol.~9, p.~e1546, 11 2023.

\bibitem{crow2002case}
R.~Crow, ``{The case for institutional repositories: a SPARC position paper},'' 2006.

\bibitem{hata2021science}
H.~Hata, J.~L. Guo, R.~G. Kula, and C.~Treude, ``Science-software linkage: the challenges of traceability between scientific knowledge and software artifacts,'' {\em arXiv preprint arXiv:2104.05891}, 2021.

\bibitem{FAIRrs}
M.~Barker, N.~{Chue Hong}, D.~Katz, A.~Lamprecht, C.~Martinez-Ortiz, F.~Psomopoulos, J.~Harrow, L.~Castro, M.~Gruenpeter, P.~Martinez, and T.~Honeyman, ``Introducing the fair principles for research software,'' {\em Scientific Data}, vol.~9, Dec. 2022.
\newblock Publisher Copyright: {\textcopyright} 2022, The Author(s).

\bibitem{garijo_2024}
D.~Garijo, M.~Arroyo, E.~Gonzalez, C.~Treude, and N.~Tarocco, ``Bidirectional paper-repository tracing in software engineering,'' in {\em Proceedings of the 21st International Conference on Mining Software Repositories}, MSR '24, (New York, NY, USA), p.~642–646, Association for Computing Machinery, 2024.

\bibitem{katz2024special}
D.~S. Katz and N.~P.~C. Hong, ``Special issue on software citation, indexing, and discoverability,'' {\em PeerJ Computer Science}, vol.~10, 2024.

\bibitem{10.1007/978-3-031-16802-4_15}
E.~Escamilla, M.~Klein, T.~Cooper, V.~Rampin, M.~C. Weigle, and M.~L. Nelson, ``The rise of github in scholarly publications,'' in {\em Linking Theory and Practice of Digital Libraries} (G.~Silvello, O.~Corcho, P.~Manghi, G.~M. Di~Nunzio, K.~Golub, N.~Ferro, and A.~Poggi, eds.), (Cham), pp.~187--200, Springer International Publishing, 2022.

\bibitem{EscamillaEmily2023CBNA}
E.~Escamilla, M.~Klein, T.~Cooper, V.~Rampin, M.~C. Weigle, M.~L. Nelson, S.~Tuarob, D.~H. Goh, and S.-J. Chen, ``Cited but not archived: Analyzing the status of code references in scholarly articles,'' in {\em Leveraging Generative Intelligence in Digital Libraries: Towards Human-Machine Collaboration}, vol.~14458 of {\em Lecture Notes in Computer Science}, pp.~194--207, Singapore: Springer Singapore Pte. Limited, 2023.

\bibitem{WATTANAKRIENGKRAI2022111117}
S.~Wattanakriengkrai, B.~Chinthanet, H.~Hata, R.~G. Kula, C.~Treude, J.~Guo, and K.~Matsumoto, ``Github repositories with links to academic papers: Public access, traceability, and evolution,'' {\em Journal of Systems and Software}, vol.~183, p.~111117, 2022.

\bibitem{10.1145/3383583.3398578}
M.~F\"{a}rber, ``Analyzing the github repositories of research papers,'' in {\em Proceedings of the ACM/IEEE Joint Conference on Digital Libraries in 2020}, JCDL '20, (New York, NY, USA), p.~491–492, Association for Computing Machinery, 2020.

\bibitem{hasselbring2019fairopencomputerscience}
W.~Hasselbring, L.~Carr, S.~Hettrick, H.~Packer, and T.~Tiropanis, ``Fair and open computer science research software,'' 2019.

\bibitem{Ukri2025Feb}
Ukri, ``{GtR},'' Feb. 2025.
\newblock [Online; accessed 25. Feb. 2025].

\bibitem{python_requests}
K.~Reitz, ``Python requests v2.32.3,'' 2024.

\bibitem{druskat2017citation}
S.~Druskat, ``Citation file format core (cff-core),'' 2017.

\bibitem{ukri_concordat}
{UKRI}, ``Concordat on open research data,'' 2016.

\bibitem{acm_repr}
ACM, ``{Artifact Review and Badging},'' Apr. 2025.
\newblock [Online; accessed 7. Apr. 2025].

\bibitem{Mens2016}
T.~Mens, ``An ecosystemic and socio-technical view on software maintenance and evolution,'' in {\em 2016 IEEE International Conference on Software Maintenance and Evolution (ICSME)}, pp.~1--8, 2016.

\bibitem{sharma2022pymafka}
A.~Sharma, ``New 'pymafka' malicious package drops cobalt strike on macos, windows, linux,'' May 2022.
\newblock Accessed: 2025-06-04.

\bibitem{kuznetsov2022lofylife}
I.~Kuznetsov and L.~Bezvershenko, ``Lofylife: malicious npm packages steal discord tokens and bank card data,'' July 2022.
\newblock Accessed: 2025-06-04.

\bibitem{checkpoint2022cloudguard}
{Check Point Research}, ``Cloudguard spectral detects several malicious packages on pypi – the official software repository for python developers,'' August 2022.
\newblock Accessed: 2025-06-04.

\end{thebibliography}
\end{document}